\documentclass[]{pasj02} 
\usepackage[switch,mathlines]{lineno} 
\usepackage{amsmath, graphicx}
\usepackage{booktabs}
\usepackage{lscape} 
\usepackage{bm}
\usepackage{comment} 

\jyear{2025}
\Received{}
\Accepted{}

\begin{document} 

\title{Spatially Resolved Plasma Diagnostics of the Supernova Remnant DEM L71 using the Reflection Grating Spectrometer}

\author{
 Yuki \textsc{Amano},\altaffilmark{1}\altemailmark
 \email{amano.yuki.t76@kyoto-u.jp} 
 Yuken \textsc{Ohshiro},\altaffilmark{2}
 Hiromasa \textsc{Suzuki},\altaffilmark{3}
 Kotaro \textsc{Fukushima},\altaffilmark{1}
 Hiroya \textsc{Yamaguchi}\altaffilmark{1, 4}
}
\altaffiltext{1}{The Institute of Space and Astronautical Science, Japan Aerospace and Exploration Agency, 3-1-1 Yoshinodai, Chuo-ku, Sagamihara 252-5210, Japan}
\altaffiltext{2}{RIKEN Pioneering Research Institute, 2-1 Hirosawa, Wako, Saitama 351-0198, Japan}
\altaffiltext{3}{Faculty of Engineering, University of Miyazaki, 1-1 Gakuen, Kibanadai Nishi, Miyazaki, Miyazaki 889-2192, Japan}
\altaffiltext{4}{Department of Physics, The University of Tokyo, 7-3-1 Hongo, Bunkyo-ku, Tokyo 113-0033, Japan}





\maketitle

\begin{abstract}
We present a spatially resolved high-resolution X-ray spectroscopy of the supernova remnant DEM L71 using the Reflection Grating Spectrometer (RGS) aboard XMM-Newton.
Because of the large dispersion angle of the RGS, we are able to resolve individual emission lines and examine their spatial distributions within this moderately extended remnant.
We derive line fluxes across different regions of DEM L71 and perform quantitative plasma diagnostics. 
Our analysis reveals that some regions have high forbidden-to-resonance ratios of O\emissiontype{VII} He$\alpha$ lines, suggesting a non-negligible contribution from additional physical processes, such as charge exchange and/or resonance scattering.
Our results demonstrate that the RGS has potential to serve as an outstanding X-ray imaging spectrometer for moderately diffuse objects.
\end{abstract}


\section{Introduction}\label{sec:introduction}
High-resolution X-ray spectroscopy of supernova remnants (SNRs) allows detailed plasma diagnostics using individual line intensities, providing insights into the physical processes in shock-heated plasmas.
For example, recent grating observations revealed the importance of charge exchange (CX) and resonance scattering (RS) in SNRs (e.g., \cite{koshiba2022, amano2020}).
These processes arise from interactions between an SNR and non-uniform surrounding gas, or from the asymmetrical structures within a remnant, emphasizing the critical role of spatially resolved high-resolution spectroscopy in understanding the diversity in the physics of SNRs.
The X-Ray Imaging and Spectroscopy Mission (XRISM \cite{tashiro2018}), launched in September 2023, is equipped with a non-dispersive high-resolution X-ray spectrometer Resolve \citep{ishisaki2022}, which enables detailed plasma diagnostics across different regions of extended sources, such as SNRs.
This capability provided new insights into the explosion mechanism of the progenitor and the formation processes of peculiar plasmas of SNRs, through constraints on the velocity structures \citep{xrism2024n132d, suzuki2025} and diagnostics of Fe K-shell lines \citep{xrism2025overionized}.
However, since the angular resolution of XRISM is aprroximately $1.3$ arcmin (half-power diameter), achieving the spatially resolved high-resolution spectroscopy is still limited.

The reflection grating spectrometer (RGS) aboard XMM-Newton provides high energy resolution ($E/\Delta E$ from 200 to 800) X-ray spectroscopy within the wavelength of 5--38 \AA~, which corresponds to 0.33--2.5 keV \citep{denherder2001}. 
In particular, the RGS has the superior energy resolution and the effective area in the energy band below 1 keV, which is suitable for plasma diagnostics using O K-shell and Fe L-shell lines (e.g., \cite{porquet2001, xu2002}).
Although the RGS is a slitless dispersive spectrometer primarily designed for point sources, its relatively large dispersion angle allows it to resolve line emissions from moderately extended sources including SNRs in the Magellanic Clouds \citep{broersen2011, rasmussen2001}.
Furthermore, as demonstrated by \cite{vanderheyden2003}, it is also possible to reconstruct the spatial distribution of emission lines from the detailed profiles in RGS spectra.
These capabilities highlight the potential of the RGS to conduct spatially resolved high-resolution spectroscopy for moderately extended objects.

\begin{figure}
\begin{center}
 \includegraphics[width=8.0cm]{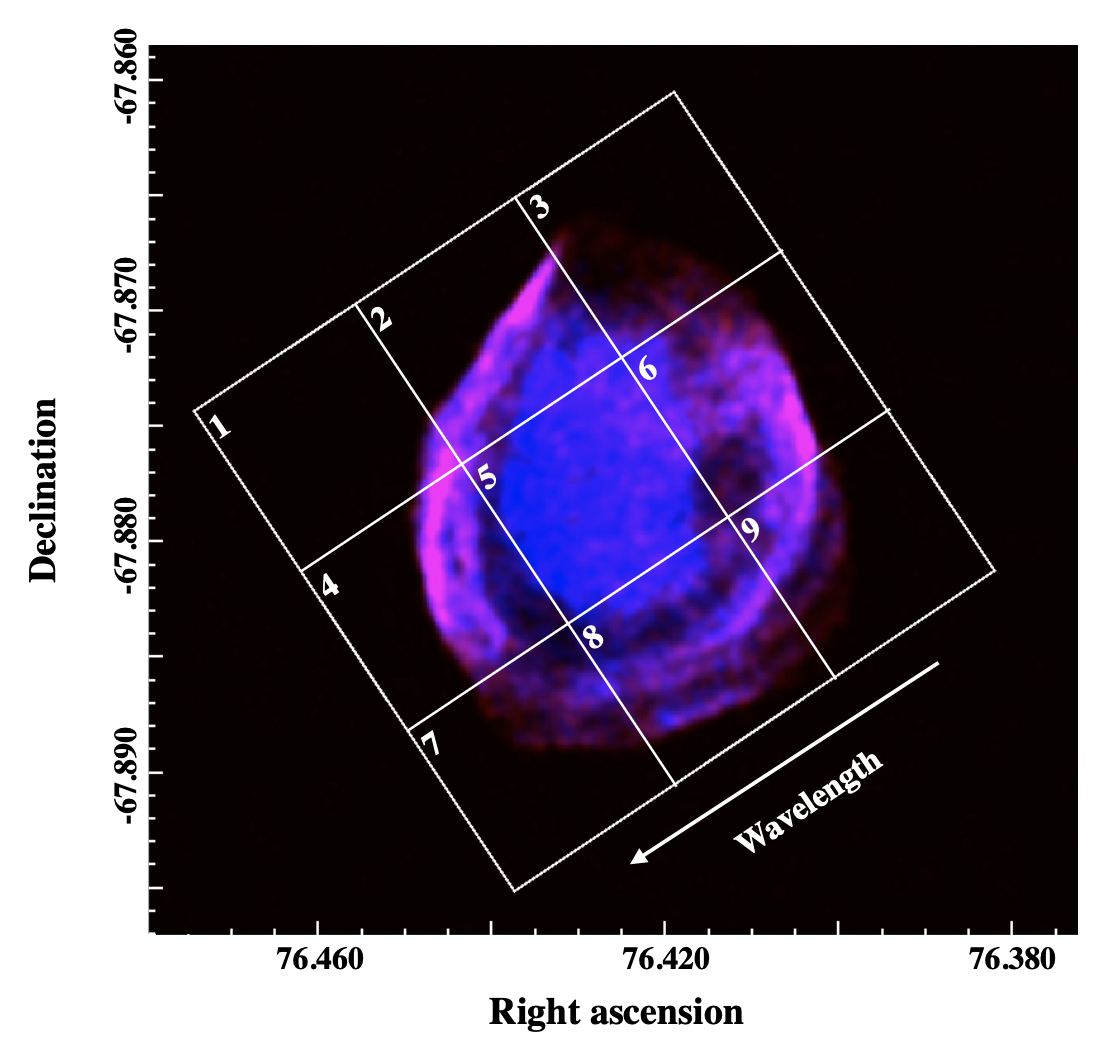} 
\end{center}
\caption{Chandra ACIS image of DEM L71: red represents the 18--21 \AA ~(0.60--0.68 keV) range, corresponding to the O\emissiontype{VIII} Ly$\alpha$ lines, while blue denotes the 12--17 \AA ~(0.75 --1.0 keV) range, associated with Fe\emissiontype{XVII} L$\alpha$ lines.
We calculate line fluxes for each region bounded by solid white lines as discribed in Section \ref{subsec:flux_ratio_map}.
The RGS dispersion direction is indicated by the white arrow.
{Alt text: Chandra image covering the RGS field-of-views of DEM L71.}
}
\label{fig:ccd_image}
\end{figure}

DEM L71 is a middle-aged ($\sim$4000 yr; \cite{ghavamian2003}) SNR located in the Large Magellanic Cloud (LMC).
Its X-ray emission is characterized by a mixture of metal-rich ejecta and a shock-heated interstellar medium (ISM) (e.g., \cite{uchida2015}).
The enhancement of Fe abundance of ejecta indicates that DEM L71 originated from a type Ia SNe (e.g., \cite{hughes1998, siegel2020}).
Chandra observations spatially resolved the central bright region of Fe-rich ejecta from the outer shell dominated by the swept-up ISM (\cite{hughes2003, alan2022}, and see also Figure \ref{fig:ccd_image} in this paper).
A previous study with the RGS revealed a difference in the line profiles of the forbidden ($f$) and resonance ($r$) of O\emissiontype{VII} He$\alpha$ lines \citep{vanderheyden2003}.
They argued that this difference reflects the spatial variations of these emission line intensities, which implies the possibility that CX and/or RS take place in the regions where the enhanced $f/r$ ratio is observed.
A more quantitative study based on spatially resolved plasma diagnostics is necessary to determine whether these processes are responsible.

Here, we present a spatially resolved high-resolution X-ray spectroscopy of DEM L71 with the RGS. 
The RGS images provide information on the spatial distribution of emission lines and the Doppler velocity of the ejecta and ISM plasmas.
Based on the detected wavelength of each line emission, we extract spatial information of emission lines along the dispersion direction.
We calculate line fluxes across different regions by taking into account foreground absorption, detector efficiency, and contribution of continuum emission.
Our method enables quantitative region-by-region plasma diagnostics, which has been considered to be difficult to achieve with grating spectrometers.
The obtained $f$/$r$ ratios of O\emissiontype{VII} He$\alpha$ of several regions imply a non-negligible contribution of additional physical processes such as CX and/or RS.  
Throughout the paper, errors are given at a 1 $\sigma$ confidence level. 
We assume the distance to DEM L71 (LMC) to be 50 kpc \citep{pietrzynski2013}.

\section{Observation and data reduction} \label{sec:observation}

We observed DEM L71 with the XMM-Newton satellite \citep{jansen2001} in April 2022 (Obs. ID 0884620101).
Besides this observation, DEM L71 was observed by XMM-Newton in 2001 and 2003.
The observation in 2003 was performed with a roll angle that aligned DEM L71 and the nearby SNR N23 along the dispersion direction of the RGS, complicating the extraction of RGS spectra purely of DEM L71.
However, the 2001 observational data are of relatively low statistical quality, so they provide no additional information.
Consequently, our analysis focused exclusively on the new observational dataset.
For spectral and imaging analysis, we used the RGS \citep{denherder2001} and European Photon Imaging Camera (EPIC) MOS \citep{turner2001} data.
Data reduction was performed with XMM-Newton Science Analysis Software (SAS) version 20.0.0.
We processed the RGS data with the RGS pipeline tool {\tt rgsproc}.
We applied good time intervals to exclude periods of background flares.
This filtering was based on the count rate in CCD number 9, which is located nearest to the optical axis of the mirror and is most susceptible to background contamination. 
The effective exposure is $\sim 130~{\rm ks}$ for both the RGS 1 and RGS 2, respectively.

\section{Analysis} \label{sec:analysis}

\subsection{RGS image} \label{subsec:monochromatic_image}

We extract RGS images to examine the spatial distribution of emission line intensities.
As the RGS is a slit-less dispersive spectrometer, RGS images contain only the spatial information of the object along the cross-dispersion axis, while along the dispersion axis, they contain both spatial and wavelength information of the incident photons. 
{According to the XMM-Newton Users Handbook,}\footnote{https://xmm-tools.cosmos.esa.int/external/xmm$\_$user$\_$support/documentation /uhb/XMM$\_$UHB.pdf} the off-axis angle $\theta$ in arcmin of the incident photon is related to the wavelength deviation $\lambda$ in \AA ~as follows:
\begin{equation}
\label{eq:RGS_LSF}
 \lambda = 0.138 ~\theta/m,
\end{equation}
where $m$ represents the spectral order.
Therefore, based on Equation (\ref{eq:RGS_LSF}), the 1.5 arcmin spatial extent of DEM L71 leads to a line broadening of 0.207 \AA, which is sufficiently small to resolve prominent emission lines such as O \emissiontype{VII} He$\alpha$ $f$ (22.10 \AA) and $r$ (21.60 \AA).

\begin{figure*}
\begin{center}
 \includegraphics[width=19.0cm]{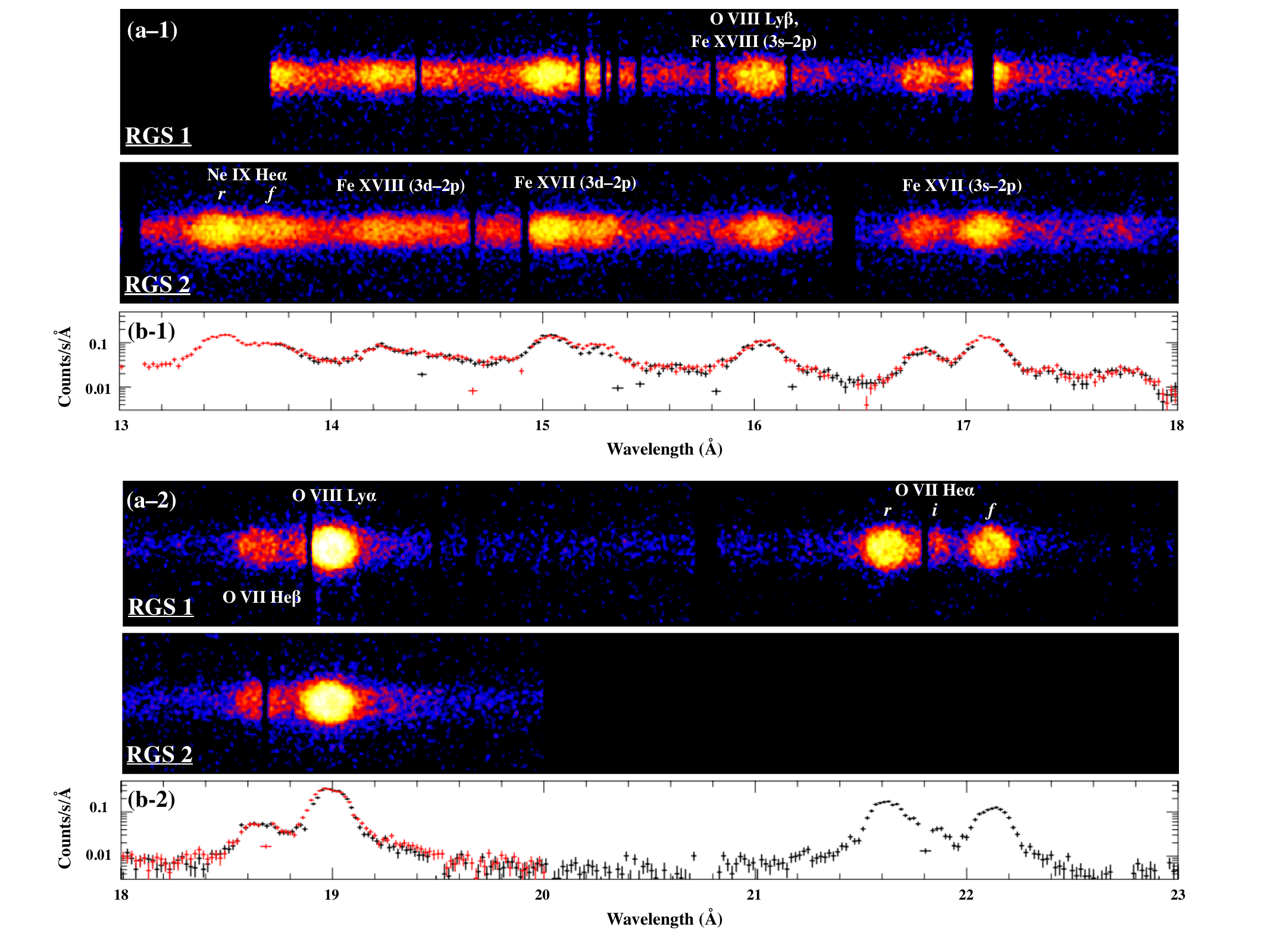} 
\end{center}
\caption{RGS images and spectra of DEM L71. 
(a) RGS images obtained with RGS 1 and 2, where the horizontal axis represents the dispersion angle and corresponding photon wavelength, and the vertical axis indicates the off-axis angle of the incident photons along the cross-dispersion axis.
(b) RGS 1 (black) and RGS 2 (red) first-order spectra.
{Alt text: 13--23 angstrom RGS images and spectra.}
}
\label{fig:RGSimage}
\end{figure*}

Figure \ref{fig:RGSimage} shows the RGS images and spectra.
We reprocess the RGS data using the SAS task {\tt evselect}.
As outlined in XMM-Newton Users Handbook, we can distinguish spectral orders by the energy resolution of the RGS focal plane CCDs.
We extract events within the pulse height range corresponding to first order photon events to produce the RGS images and spectra.
RGS 1 and 2 cover wavelength ranges of 5.2--38.2 \AA~and 4.7--37.3 \AA, which are converted to spatial fields of view of $\sim 239$ and $\sim 236$ arcmin, based on Equation (\ref{eq:RGS_LSF}), respectively.
Since the field of view along the cross-dispersion axis of the RGS is $5$ arcmin, the spatial distribution of each emission line is reconstructed by setting the aspect ratio of the RGS 1 and 2 images to 239:5 and 236:5, respectively.

\begin{figure*}
\begin{center}
 \includegraphics[width=18.0cm]{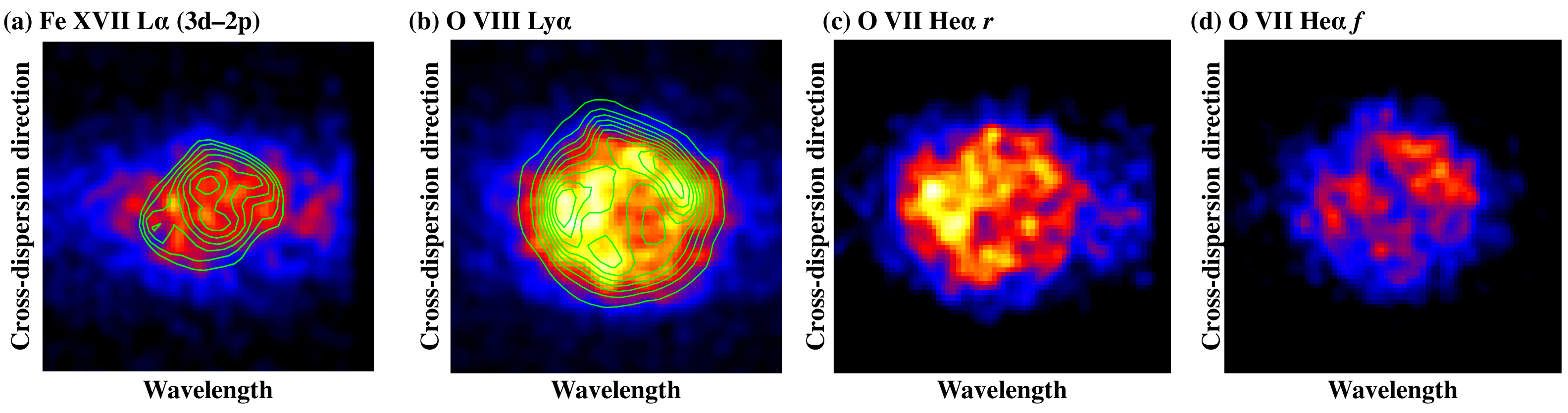} 
\end{center}
\caption{Close-up views of the RGS images around the (a) Fe\emissiontype{XVII} L$\alpha$ (3d--2p), (b) O\emissiontype{VIII} Ly$\alpha$, (c) O\emissiontype{VII} He$\alpha$ resonance and (d) O\emissiontype{VII} He$\alpha$ forbidden. 
The solid contours in (a) and (b) represent the EPIC MOS band images corresponding to O\emissiontype{VIII} Ly$\alpha$ (18--21 \AA ~or 0.60--0.68 keV) and Fe\emissiontype{XVII} L$\alpha$ (12--17 \AA ~or 0.75--1.0 keV), respectively. 
{Alt text: Close-up views of the RGS images. 
The X-axis corresponds to the wavelength, and the Y-axis corresponds to the off-axis angle along the cross-dispersion direction.}
}
\label{fig:RGS_image_focus}
\end{figure*}

\begin{figure*}
\begin{center}
 \includegraphics[width=19.0cm]{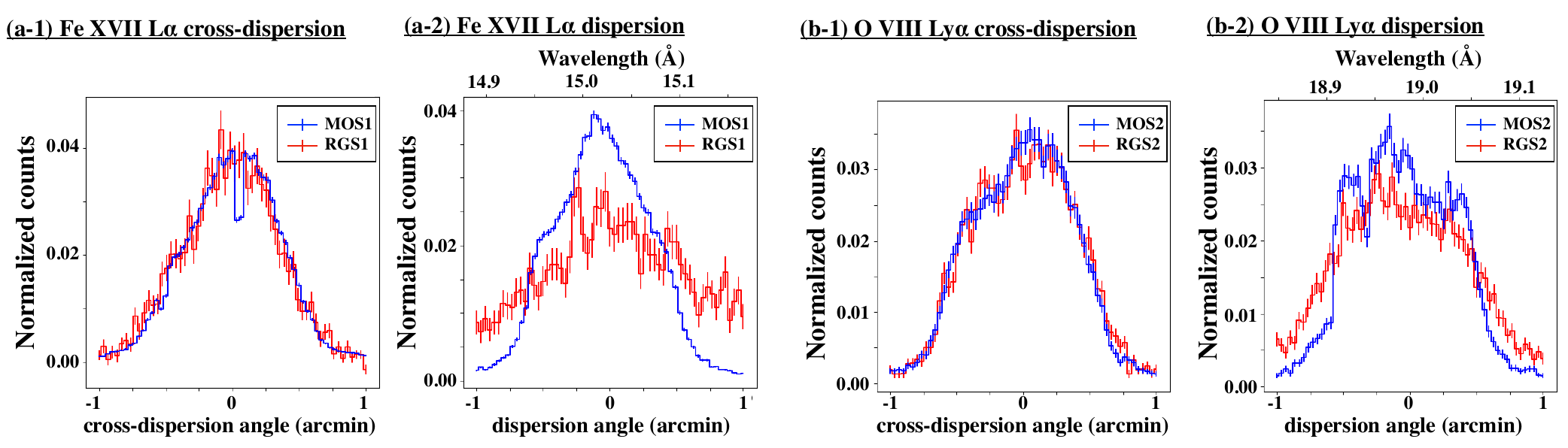} 
\end{center}
\caption{Comparison of line profiles obtained with the RGS and MOS.
(a) Cross-dispersion and dispersion profiles of the Fe \emissiontype{XVII} L$\alpha$ (3d--2p) line at 15 \AA. 
The count number for each bin is normalized by the total count number.
MOS profiles are created with X-ray events in the range of 12--17 \AA ~or 0.75--1.0 keV.
(b) Same as (a) but for profiles of the O\emissiontype{VIII} Ly$\alpha$ lines.
MOS profile are created with X-ray events in the range of 18--21 \AA ~or 0.60--0.68 keV.
{Alt text: Eight line graphs.}
}
\label{fig:broadening}
\end{figure*}

We derive several implications from the RGS images of Fe and O lines. 
Figures \ref{fig:RGS_image_focus} (a) and (b) provide close-up views of the RGS images around O\emissiontype{VIII} Ly$\alpha$ and Fe\emissiontype{XVII} L$\alpha$ (3d--2p) lines.
The solid contours correspond to MOS band images around the O\emissiontype{VIII} Ly$\alpha$ and Fe\emissiontype{XVII} L$\alpha$ (3d--2p) lines. 
The Fe\emissiontype{XVII} L$\alpha$ (3d--2p) line exhibits a centrally filled morphology, while O\emissiontype{VIII} Ly$\alpha$ shows a limb-brightened structure.
These morphologies suggest that O and Fe emissions originate from swift-up ISM and ejecta, respectively, consistent with the results of the Chandra observation \citep{hughes2003}.
Additionally, we find that these emission lines exhibit broader profiles along the dispersion axis than those expected from the spatial extent of DEM L71.
Figure \ref{fig:broadening} shows comparison of the RGS and MOS line profiles.
We subtract background and continuum emission from the RGS line profiles.
The contribution of the background and continuum emission are estimated based on the flux in the 19.6–20.0 \AA.
While the RGS and MOS profiles exhibit similar extents along the cross-dispersion axis as shown in Figure \ref{fig:broadening} (a-1) and (b-1) , the RGS appears broader along the dispersion axis than the MOS profiles (Figure \ref{fig:broadening} (a-2) and (b-2)).
We attribute these broadening to Doppler effects caused by the expansion of the ISM shell and the ejecta. 
The broadening of the O\emissiontype{VIII} Ly$\alpha$ lines is relatively slight, whereas the broadening of the  Fe\emissiontype{XVII} L$\alpha$ (3d--2p) line is significant, which can be interpreted as reflecting differences in the expansion velocity of the ISM shell and ejecta.
Detailed measurements and interpretations of the expansion velocity will be presented in a subsequent publication.
Furthermore, comparing (c) and (d) panels in Figure \ref{fig:RGS_image_focus}, we observe distinct spatial distributions for the O\emissiontype{VII} He$\alpha$ $f$ and $r$ lines.
In this study, we focus on this spatial variation between these lines.

\subsection{Flux ratio map} \label{subsec:flux_ratio_map}

\begin{figure*}
\begin{center}
 \includegraphics[width=18.0cm]{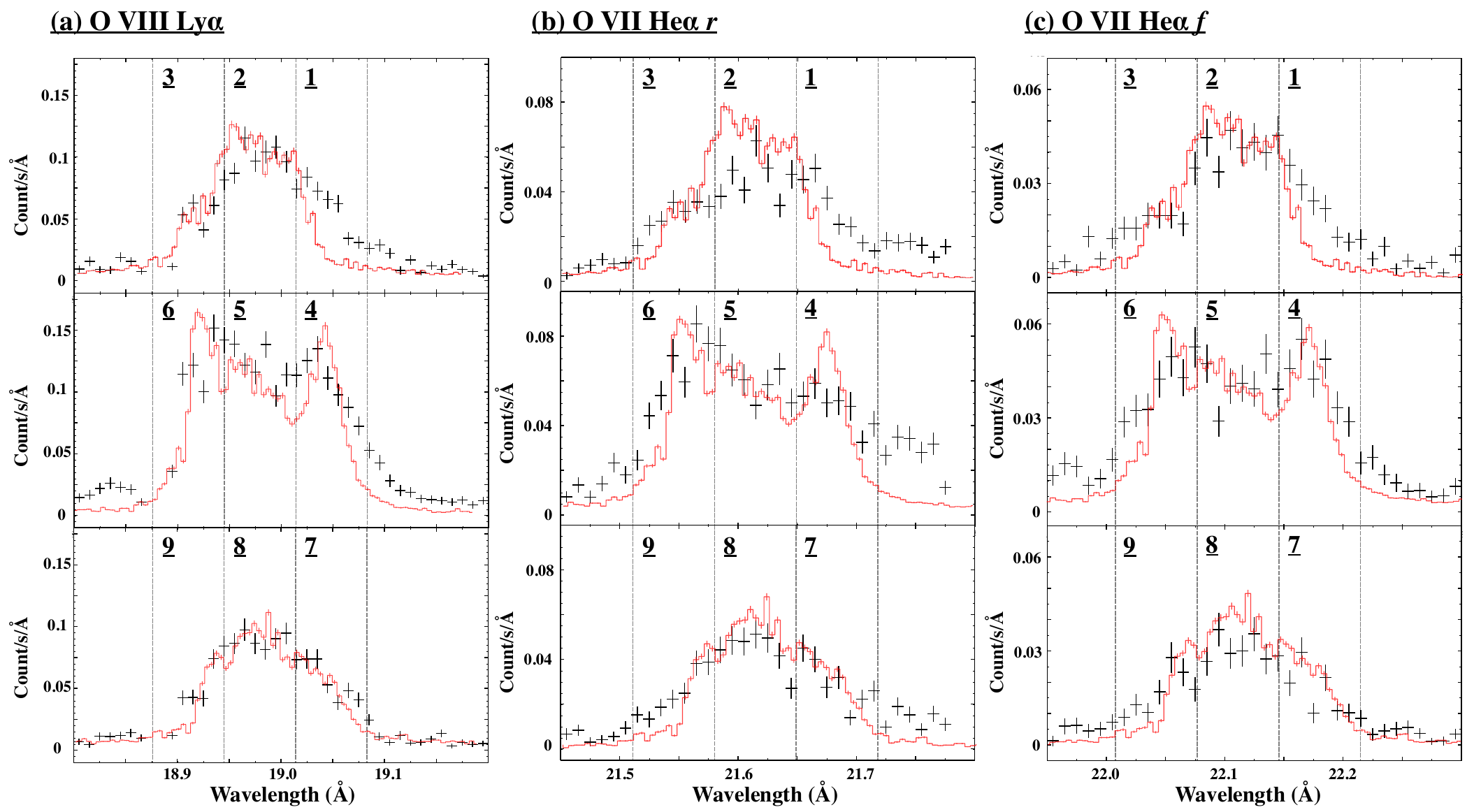} 
\end{center}
\caption{RGS spectra of (a) O\emissiontype{VIII} Ly$\alpha$, (b) O\emissiontype{VII} He$\alpha$ resonance and (c) O\emissiontype{VII} He$\alpha$ forbidden lines.
The top panels present spectra extracted from the segment corresponding to the regions 1--3 in Figure \ref{fig:ccd_image}, the middle panels correspond to the regions 4--6, and the bottom panels correspond to the regions 7--9.
The MOS2 profile for the region corresponding to each segmentation is shown by a solid red line.
We create the MOS profile with X-ray events corresponding to the O\emissiontype{VIII} Ly$\alpha$ lines (18--21 \AA ~or 0.60--0.68 keV).
The segmentation of the emission region along the dispersion axis for each line is determined using Equation (\ref{eq:RGS_LSF}), with the boundaries of each region marked by gray dashed lines. 
The region numbers are labeled consistently with those in Figure \ref{fig:ccd_image}.
{Alt text: RGS spectrum for each region.}
}
\label{fig:region_spectra}
\end{figure*}

We calculate the fluxes of O\emissiontype{VIII} Ly$\alpha$, O\emissiontype{VII} He$\alpha$ $r$, and O\emissiontype{VII} He$\alpha$ $f$ lines in the 9 regions shown in Figure \ref{fig:ccd_image} to perform quantitative diagnostics.
First, the source region is divided into three segments, each with a width of $0.5$ arcmin along the cross-dispersion axis.
Figure \ref{fig:region_spectra} shows close-up view of the background-subtracted spectra of each of the three segments, where the background spectrum is extracted from an off-source region in $1.5$ arcmin along the cross-dispersion axis.
According to Equation (\ref{eq:RGS_LSF}), the emission regions of the individual lines are spatially resolved along the dispersion axis based on the observed photon wavelengths.
The intrinsic wavelengths of each line are defined by applying a systematic red shift of $260~{\rm km~s}^{-1}$ due to radial motion of the LMC \citep{vandermarel2002} to the wavelengths in the rest frame obtained by referring to AtomDB \citep{smith2001, foster2012}.
The relationships between the observed wavelengths and the corresponding emission regions are indicated as gray dashed lines in Figure \ref{fig:region_spectra}.
The flux for each region is calculated by correcting the observed count rate for the effective area of the RGS and foreground absorption. 
The foreground absorption for Galactic and LMC components are calculated using the {\tt tbnew} models.
The hydrogen column density of the former is fixed to $6~\times~10^{20}~{\rm cm}^{-2}$ \citep{dickey1990} whereas that of the latter is fixed to $3.7~\times~10^{21}~{\rm cm}^{-2}$ \citep{uchida2015}.
The elemental abundances for the Galactic and LMC absorption are fixed to solar values \citep{wilms2000} and LMC values reported by previous studies ($\sim$~0.3 solar; \cite{russell1992, schenck2016}).
The line fluxes are obtained by subtracting the flux of the continuum component from the fluxes in each region.
The contribution of the continuum emission is estimated based on the flux in the 20.0--20.6 \AA, where we confirm that there is no major emission line.

\begin{figure*}
\begin{center}
 \includegraphics[width=18.0cm]{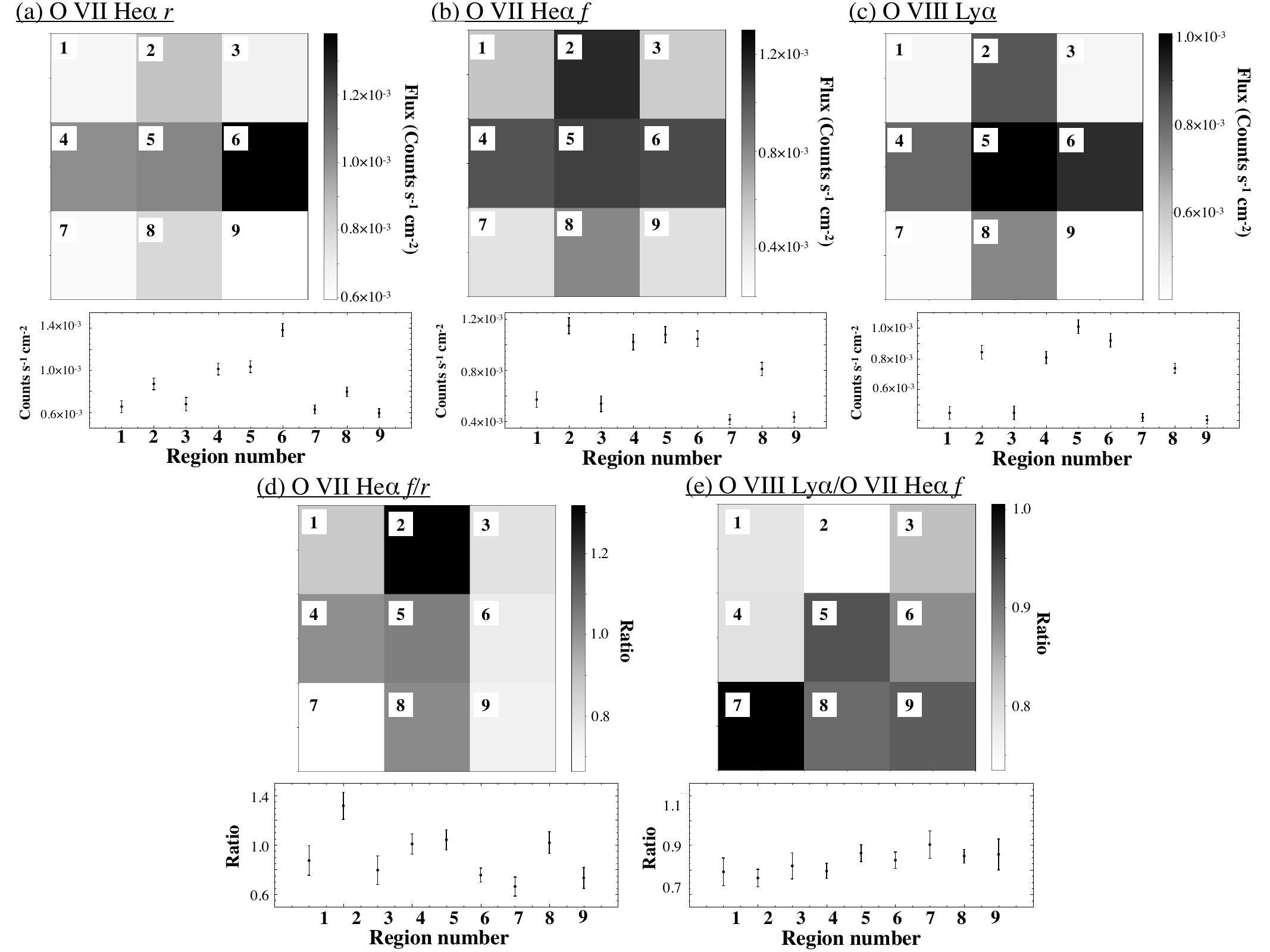} 
\end{center}
\caption{Fluxes and their ratio maps for (a) the O\emissiontype{VII} He$\alpha$ resonance line flux, (b) the O\emissiontype{VII} He$\alpha$ forbidden flux, (c) the O\emissiontype{VIII} Ly$\alpha$ lines flux, (d) the O\emissiontype{VII} He$\alpha$ $f/r$ ratio, (e) the O\emissiontype{VIII} Ly$\alpha$/O\emissiontype{VII} He$\alpha$ $f$ ratio. 
The upper panels display maps of the fluxes and ratios, with region numbers consistent with those shown in Figure \ref{fig:ccd_image}. 
The lower panels present corresponding plots of these quantities. 
{Alt text: Black and white scale maps and scattered plots.}
}
\label{fig:flux_map}
\end{figure*}

The line fluxes and their ratios in each region are presented in Figure \ref{fig:flux_map}, where our quantitative analysis reveals spatial variations among these fluxes and ratios. 
As seen in Figures \ref{fig:flux_map} (b) and (c), O\emissiontype{VIII} Ly$\alpha$ and O\emissiontype{VII} He$\alpha$ $f$ exhibit similar spatial distributions.
Indeed, the O\emissiontype{VIII} Ly$\alpha$/O\emissiontype{VII} He$\alpha$ $f$ ratio shows a uniform distribution at approximately 0.9 across all regions (Figure \ref{fig:flux_map} (e)).
On the other hand, the spatial distribution of the O\emissiontype{VII} He$\alpha$ $r$ line deviates from the distribution of these lines.
For example, in the middle segment (the regions 4--6), the fluxes of O\emissiontype{VII} He$\alpha$ $r$ differ between regions 4 and 6. 
This $r$ line distribution leads to an enhanced $f/r$ ratio in the region 4, as seen in Figure \ref{fig:flux_map} (d).
This trend is consistent with the enhancement of the $f/r$ ratio in the eastern region previously reported by \citet{vanderheyden2003}.
Furthermore, we find high $f/r$ ratios exceeding unity not only in region 4 but also in regions 2, 5, and 8.

\subsection{Line diagnostics with collisionally ionized plasma model} \label{subsec:diagnostics_NEI}

\begin{figure*}
\begin{center}
\includegraphics[width=17.0cm]{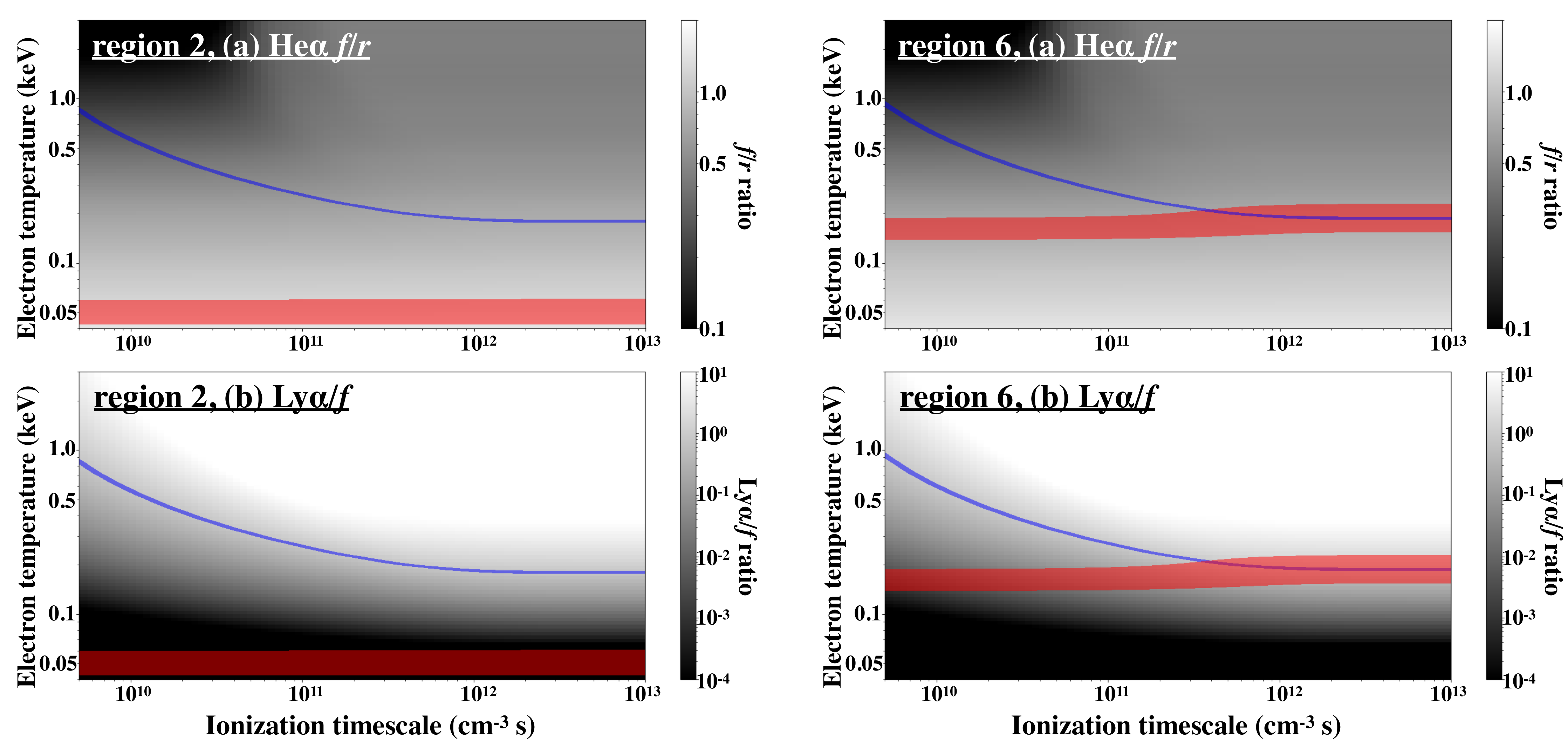} 
\end{center}
\caption{Comparison of the observed O line intensity ratios and those predicted by plasma models.
Theoretical predictions of AtomDB plasma models for (a) O He$\alpha$ $f$/$r$, (b) O Ly$\alpha$/He$\alpha$ $f$ as a function of ionization time scale ($n_{\rm e}t$: horizontal axis) and electron temperature ($kT_{\rm e}$: vertical axis) are shown in the upper and lower panels, respectively. 
The observed intensity ratios and their 1$\sigma$ uncertainty ranges of He$\alpha$ $f$/$r$ and Ly$\alpha$/He$\alpha$ $f$ are indicated with red and blue hatched regions.
The left and right panels show observational values for regions 2 and 6, respectively.
{Alt text: Four black and white scale maps whose x-axis show ionization timescale, and y-axis shows electron temperature.}
}
\label{fig:kTe_net_map}
\end{figure*}

We conduct plasma diagnostics to investigate the cause of the spatial variation in the $f$/$r$ ratios.
In this analysis, we attribute the O line emissions from DEM L71 to the shocked ISM, based on their shell-like morphology observed in the RGS images (Figure \ref{fig:RGS_image_focus} (b)).
Figure \ref{fig:kTe_net_map} presents a comparison between the observed $f$/$r$ and Ly$\alpha$/$f$ ratios for each region and theoretical predictions derived from a collisionally ionized plasma model.
The theoretical ratios are calculated as functions of electron temperature ($kT_{\rm e}$) and ionization timescale ($n_{\rm e}t$) using {\tt PyAtomDB}.
The observed $f$/$r$ and Ly$\alpha$/$f$ ratios are shown as red and blue hatched regions, respectively.
Here we mainly explain the results obtained in regions 2 and 6. 
Results for the other regions are summarized in the Appendix.

We first focus on the region 6, where the observed $f$/$r$ ratio is relatively small.
The red hatched region in the region 6 (a) in Figure \ref{fig:kTe_net_map} represents the observed $f$/$r$ ratio and the corresponding range of the $kT_{\rm e}$ and $n_{\rm e}t$.
The $f/r$ ratio primarily depends on $kT_{\rm e}$, and the observed $f$/$r$ value requires a $kT_{\rm e}$ range of 0.1--0.3 keV, regardless of the value of $n_{\rm e}t$.
On the other hand, the blue hatched region in the region 6 (b) in Figure \ref{fig:kTe_net_map} shows the observed Ly$\alpha$/$f$ ratio, which reflects the relative populations of H-like and He-like ions and provides a constraint on $n_{\rm e}t$. 
The overlap of the red and blue hatched regions indicates that these O lines originate from a plasma with $kT_{\rm e} \approx 0.2~{\rm keV}$ and $n_{\rm e}t > 10^{12}~{\rm cm^{-3}s}$ (i.e., collisional ionization equilibrium state).
We obtain similar results in regions 1, 3, 7, and 9, suggesting that the ISM plasmas of DEM L71 in these regions consist of collisional ionization equilibrium plasmas with $kT_{\rm e} \approx 0.2~{\rm keV}$.
This result is consistent with previous spectral analyses for the entire region of DEM L71 using RGS and Suzaku \citep{uchida2015, vanderheyden2003}.

Next, we focus on the region 2, which has the largest $f/r$ ratio.
In this region, the $f$/$r$ ratio requires a lower $kT_{\rm e}$ of 0.04--0.06 keV, as shown in the region 2 (a) in Figure \ref{fig:kTe_net_map}.
As shown in the region 2 (b) Figure \ref{fig:kTe_net_map}, such a low $kT_{\rm e}$ value predicts smaller Ly$\alpha$/$f$ values than the observed values.
This is because oxygen in such plasma is mainly in the ${\rm O}^{6+}$ state with few ${\rm O}^{7+}$ ions (e.g., \cite{bryans2006, vink2012}).
Therefore, we conclude that the observed O line ratios cannot be consistently explained by a single NEI plasma model due to their high $f$/$r$ ratios.
We obtain the same results in regions 2, 4, 5, and 8.
These discrepancies could be explained by the multiple-temperature plasmas.
However, it is unlikely that such low-temperature plasma is the major plasma component in these regions of DEM L71 since a previous region-by-region spectral analysis with Chandra has not detected any evidence of such low-temperature plasma in these regions \citep{alan2022}.
Therefore, it is worth discussing the possibility that the observed $f$/$r$ ratios are modified by other effects, such as RS and CX.

\section{Discussion} \label{sec:discussion}

We performed spatially resolved high-resolution X-ray spectroscopy of LMC SNR DEM L71 utilizing the RGS.
The wavelength resolution of RGS is approximately 0.075 \AA, which corresponds to the spatial resolution in the dispersion direction of 0.5 arcmin.
This spatial resolution makes it possible to perform quantitative region-by-region plasma diagnostics for moderately diffused objects which are difficult to be resolved even with the XRISM.
As a result, we found anomalously high $f/r$ ratios in some regions, which imply possible contributions of CX and/or RS. 
Here we perform diagnostics to investigate whether these processes explain the observed line ratios.

\subsection{Resonance scattering} \label{sec:RS}

RS is an apparent scattering phenomenon due to an absorption and subsequent reemission of line photons by ions.
The RS effect can reduce intensities of resonance lines, inducing the enhancement of the $f/r$ ratio.
The RS effect is generally negligible in SNRs, however, \cite{kaastra1995} pointed out that RS can occur in a plasma with a large depth along the line of sight such as a rim of SNRs.
Indeed, \citet{amano2020} obtained observational evidence for RS from recent grating observations of LMC SNR N49.
As shown in the middle panel of Figure \ref{fig:region_spectra} (b), the difference in the $f/r$ ratio between the regions 4 and 6 of DEM L71 appears to be primarily due to the variation in the resonance line fluxes, which may be explained by the effect of RS.

We perform a similar diagnostic as described in Section \ref{subsec:diagnostics_NEI}, using theoretically predicted line ratios modified by the RS effect.
We parameterize the RS effect as the hydrogen column density $N_{\rm H}$ and investigate a range of $N_{\rm H}$, which consistently explains the observed O\emissiontype{VII} $f$/$r$ and O\emissiontype{VIII} Ly$\alpha$/O\emissiontype{VII} He$\alpha$ $f$ ratios.
Following the methodology of \cite{kaastra1995}, we calculate transmission factors for each line, assuming that a photon completely escapes from the line of sight at every scattering event. 
The transmission factor is approximated as
\begin{equation}
\label{eq:transmission}
T = \frac{1}{1 + 0.43 \tau},
\end{equation}
where $\tau$ is the optical depth at the line centroid \citep{kastner1990}.
The optical depth $\tau$ is given
\begin{equation}
\label{eq:RS_tau}
 \tau = \frac{4.24 \times 10^{6} f N_{\rm H} \left(\frac{n_{\rm i}} {n_{\rm z}}\right) \left(\frac{n_{\rm z}}{n_{\rm H}}\right) \left(\frac{M}{T_{\rm keV}}\right)^{1/2}}{E_{\rm eV}\left(1 + \frac{0.0522 M v_{100}^2}{T_{\rm keV}}\right)^{1/2}},
\end{equation}
where $f$ is the oscillator strength of the line, $E_{\rm eV}$ is the line centroid energy in eV, $N_{\rm H}$ is the hydrogen column density in ${\rm cm^{-2}}$, $n_{\rm i}$ is the number density of the ion, $n_{\rm Z}$ is the number density of the element, $M$ is the atomic weight of the ion, $T_{\rm keV}$ is the ion temperature in keV, and $v_{100}$ is the micro-turbulence velocity in units of $100~{\rm km}~{\rm s}^{-1}$.
We take the oscillator strengths from AtomDB and assume the abundance of oxygen to be the ISM value of the LMC ($\sim 0.3~{\rm solar}$; \cite{russell1992, wilms2000}).
We assume a thermal equilibrium between all ions and electrons and neglect the micro-turbulence velocity.
Under these assumptions, the optical depth $\tau$ depends on the hydrogen column density $N_{\rm H}$, $kT_{\rm e}$, and $n_{\rm e}t$ of the absorber plasma.
We obtain the ion fraction ($n_{\rm i}/n_{\rm z}$) of oxygen at each $kT_{\rm e}$ and $n_{\rm e}t$ value with {\tt PyAtomDB}.
We calculate theoretical line ratios modified by the RS effect by multiplying the line emissivities with the corresponding transmission factors calculated for each $kT_{\rm e}$ and $n_{\rm e}t$ value.

\begin{figure*}
\begin{center}
 \includegraphics[width=18.0cm]{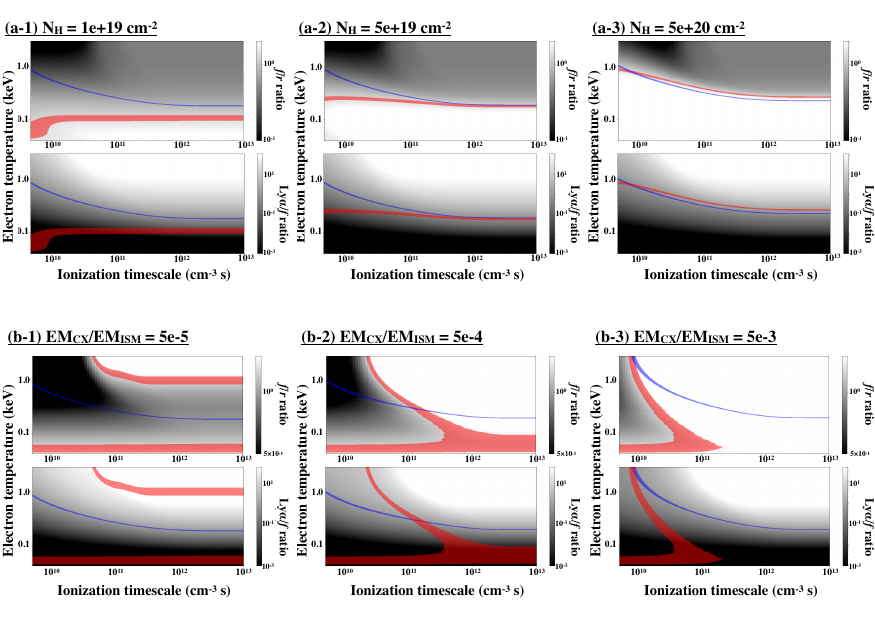} 
\end{center}
\caption{Comparison of the observed line ratios and those predicted by plasma models that takes contribution of CX and RS into consideration.
(a) Comparison with theoretical prediction modified by the RS effect of O He$\alpha$ $f$/$r$ (upper panels) and O Ly$\alpha$/He$\alpha$ $f$ (lower panels), with cases for (a–1) $N_{\rm H} = 1.0 \times 10^{19}$, (a–2) $N_{\rm H} = 5.0 \times 10^{19}$ and (a–3) $N_{\rm H} = 5.0 \times 10^{20}~{\rm cm^{-2}}$.
The observed ratios and their 1$\sigma$ uncertainty ranges of He$\alpha$ $f$/$r$ and Ly$\alpha$/He$\alpha$ $f$ in the region 2 are indicated with red and blue regions.
(b) Comparison with ``NEI $+$ CX'' model, with cases for (b–1) $EM_{\rm CX}/EM_{\rm ISM} = 5.0 \times 10^{-5}$, (b–2) $EM_{\rm CX}/EM_{\rm ISM} = 5.0 \times 10^{-4}$, (b–3) $EM_{\rm CX}/EM_{\rm ISM} = 5.0 \times 10^{-3}$.
{Alt text: Twelve black and white scale maps whose x-axis show ionization timescale, and y-axis shows electron temperature.}
}
\label{fig:RSCX_map}
\end{figure*}

Figure \ref{fig:RSCX_map} shows examples of line ratios modified by the RS effect, with cases for $N_{\rm H} = 1.0 \times 10^{19}, 5.0 \times 10^{19},  5.0 \times 10^{20}~{\rm cm^{-2}}$.
The red and blue contours represent the $f/r$ and Ly$\alpha$/He$\alpha$ $f$ ratios observed in the region 2, where the largest $f/r$ ratio is observed.
When $N_{\rm H} = 5.0 \times 10^{19}~{\rm cm^{-2}}$, the red and blue hatched regions overlap when $kT_{\rm e} \sim 0.3$~keV  (Figure \ref{fig:RSCX_map} (a–2)).
Assuming $kT_{\rm e}$ of the absorber to be below 0.9 keV, typical for the ISM of DEM L71 \citep{alan2022}, we confirm that the line ratios observed in the region 2 can be explained when $N_{\rm H}$ is in the ranges $(5.0$--$10)\times10^{19}~{\rm cm^{-2}}$.
This column density corresponds to a line-of-sight length range of 16--32 $\times (n_{\rm H}~{\rm cm^3})~{\rm pc}$. 
These plasma depths are comparable to the diameter of DEM L71 ($\sim 22~{\rm pc}$).
The results for all regions are summarized in Table \ref{tab:line_flux}.
The line ratios observed in other regions can also be explained by the effect of RS due to the plasma with column density of the same order as the region 2.
We therefore conclude that the effect of RS can explain the observed line ratios.

\begin{figure}
\begin{center}
\includegraphics[width=8.0cm]{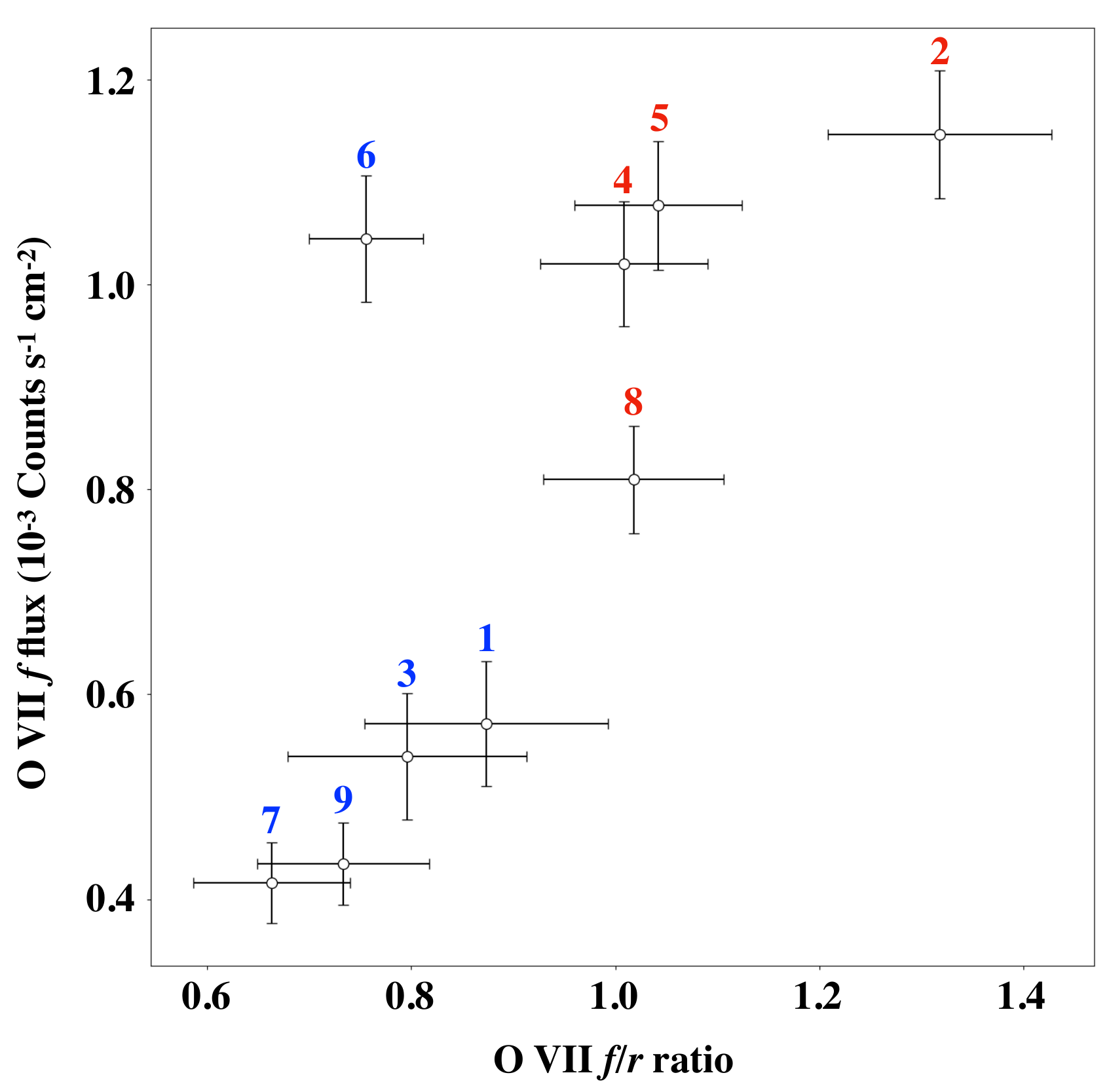} 
\end{center}
\caption{
Correlation between the O \emissiontype{VII} $f/r$ ratios and the O \emissiontype{VII} $f$ flux for each region of DEM L71.
Region numbers are marked in red and blue above each data point.
The red numbers correspond to regions where the enhanced $f/r$ ratio is observed.
{Alt text: Nine data points in each of the left and right panels.}
}

\label{fig:tau_flux}
\end{figure}

To investigate the possibility of RS, we examine correlations between the O \emissiontype{VII} $f/r$ ratio and other line fluxes and ratios.
As a result, we find a positive correlation between the $f/r$ ratio and the flux of the O \emissiontype{VII} $f$ line (Figure \ref{fig:tau_flux}), while the region 6 deviates from the correlation.
One possible interpretation of this correlation is that the RS effect primarily causes the enhanced $f/r$ ratios.
Assuming that the ISM plasma of DEM L71 has a spherical-shell morphology and a uniform electron temperature, the O \emissiontype{VII} $f$ line flux reflects the plasma density.
Therefore, regions with higher $f$ flux are expected to have larger column densities.
If the optical depth of the O \emissiontype{VII} $r$ line is close to unity, the $f/r$ ratio increases with increasing column density, which may explain the observed positive correlation.
To estimate the optical depth of each region of DEM L71, we assume a spherical shell for the ISM plasma with inner and outer radii of 0.50 and 0.55 arcmin from the Chandra X-ray image, and a uniform temperature of 0.2 keV as a typical value obtained by our line diagnostics performed in Section \ref{subsec:diagnostics_NEI}.
We derive the emission measure from the O \emissiontype{VII} $f$ line flux, estimate the plasma density in each region, and then calculate the optical depth using Equation (\ref{eq:RS_tau}).
The optical depth depends on the turbulence velocity. 
For a velocity of 250 ${\rm km} ~ {\rm s}^{-1}$, the optical depths in each region are estimated to be 0.6--1.0, indicating values close to unity and supporting the interpretation that RS contributes to the enhanced $f/r$ ratios.
Previous studies proposed the possibility of RS occurring in foreground plasma, such as the galactic halo \citep{sun2025, amano2020, gu2016}.
In this scenario, a uniform optical depth should be observed in all regions.
Although the effect of RS due to the foreground plasma cannot be rejected, the region-by-region variations in optical depth shown above support that RS occurs within DEM L71 itself.

\subsection{Charge exchange} \label{sec:CX}

CX is another possibility that causes the observed high $f$/$r$ ratio.
Observational evidence for CX was obtained from recent grating observations of middle-aged SNRs such as Puppis A \citep{katsuda2012} and the Cygnus Loop \citep{uchida2019}.
Theoretically, \cite{lallement2004} predicted that CX can be observed in the outer edge of the shock front of DEM L71.
In our analysis, the $f/r$ ratios observed in region 6 appear to deviate from the correlation seen in the other regions (Figure \ref{fig:tau_flux}).
Furthermore, in the region 5, the ISM plasma is expected to exhibit an elongated structure perpendicular to the line of sight, where RS would suppress the $f/r$ ratio by increasing the observed resonance line flux. 
Therefore, it is difficult to explain the observed high $f/r$ ratio in the region 5 by RS.
Since CX can make the $f/r$ ratio higher regardless of the plasma geometry, CX may provide a more suitable explanation for the observed enhancements of $f/r$ ratio.

We perform the same diagnostic as described in Section \ref{sec:RS}, using ``${\rm NEI} + {\rm CX}$'' model.
We assume that CX occurs between the ISM plasma of DEM L71 and the surrounding neutral gas.
We parameterize the CX contribution as the emission measure ratio ($EM_{\rm CX}/EM_{\rm ISM}$), where $EM_{\rm ISM}$ ($= n_{\rm e}n_{\rm p}V_{\rm ISM}$, where $n_{\rm e}$, $n_{\rm p}$, and $V_{\rm ISM}$ are the electron density, proton density, and the volume of ISM plasma) and $EM_{\rm CX}$ ($= n_{\rm p}n_{\rm NH}V_{\rm CX}$, where $n_{\rm NH}$, and $V_{\rm CX}$ are the neutral hydrogen density and the volume of the region where CX takes place) represent the emission measures of the ISM plasma and CX in DEM L71, respectively.
We obtain the line emissivities of the ``${\rm NEI} + {\rm CX}$'' model by summing the contributions from the NEI and CX models.
We calculate the line emissivity due to CX with {\tt VACX2} model version 2.1.3 from Atomdb.
The abundance ratio of (O/H) is fixed to be the ISM value of the LMC (0.3 solar; \cite{russell1992}). 
We assume the collision velocity between ion and neutral material to be $1000~{\rm km/s}$, which is the forward shock velocity obtained by optical observations \citep{ghavamian2003}. 
Recent studies suggest that CX emission is particularly enhanced in a region where a shock interacts with dense gas \citep{uchida2019, koshiba2022}.
Therefore, in regions where CX occurs efficiently, the forward shock may be locally decelerated and the collision velocity may be smaller.
However, the total cross section of CX does not change significantly in the collision velocity range of 100--1000 km/s \citep{gu2016cx}, and we confirm that there was no significant difference in our results when the collision velocity is 1000 km/s or 200 km/s.
The parameter ``acxmodel'', which determines the ($n$, $l$) distribution of the exchanged electrons, is set to default value of 8, as we confirm that this parameter does not significantly affect the results. 
We assume a multiple collision case, where an ion repeatedly undergoes CX until it is fully neutralized.

Figure \ref{fig:RSCX_map} shows observed O line ratios for region 2 and the line ratios of ``${\rm NEI} + {\rm CX}$'' model, where cases of $EM_{\rm CX}/EM_{\rm ISM} = 5.0\times10^{-5}, 5.0\times10^{-4}, 5.0\times10^{-3}$ are shown.
In the "NEI+CX" model, the NEI and CX model have common $kT_{\rm e}$ and $n_{\rm e}t$ values because we assumed that CX occurs between the ISM plasma of DEM L71 and the neutral gas.
Therefore, we calculate line emissivities of the NEI and CX  models for each $kT_{\rm e}$ and $n_{\rm e}t$ value on the horizontal and vertical axes of Figure \ref{fig:RSCX_map}.
The red and blue hatched regions overlap when $EM_{\rm CX}/EM_{\rm ISM} \sim 5.0\times10^{-4}$, indicating that the observed O line ratios can be explained by the ``${\rm NEI} + {\rm CX}$'' model.
We confirm that the O line ratios observed in region 2 can be explained when $EM_{\rm CX}/EM_{\rm ISM}$ is in the ranges $0.0002$--$0.005$.

We constrain the area where CX efficiently take place from $EM_{\rm CX}/EM_{\rm ISM}$.
Similar to \citet{tanaka2022}, we assume that each region of DEM L71 is a partial spherical shell with outer radius $r_{s(out)}$ and inner radius $r_{s(in)}$.
According to \citet{lallement2004} and \citet{tanaka2022}, the size of the CX-emitting region is quantitatively evaluated using the  ``collision parameter $p$''.
When the CX-emitting region has a thickness of $\Delta r$ from the shock front, $p$ is given by
\begin{equation}
\label{eq:col_par_p}
p = \frac{r_{s(out)} - \Delta r}{r_{s(out)}}.
\end{equation}
From Equation (2) of \citet{tanaka2022}, the ratio of the volume of the CX emitting region to the volume of ISM plasma of DEM L71 is expressed as 
\begin{equation}
\label{eq:col_par}
\frac{V_{\rm CX}}{V_{\rm ISM}} = \frac{1 - p^3}{1 - \left( \frac{r_{s(in)}}{r_{s(out)}} \right)^3}. 
\end{equation}
The ratio of the volume emission measure of the CX component to that of the ISM component is
\begin{equation}
\label{eq:EM_ratio}
\frac{EM_{\rm CX}}{EM_{\rm ISM}} = \frac{n_{\rm p}n_{\rm NH}V_{\rm CX}}{n_{\rm e}n_{\rm p}V_{\rm ISM}} \simeq \frac{n_{\rm NH}V_{\rm CX}}{1.2n_{\rm p}V_{\rm ISM}},
\end{equation}
where we assume $n_{\rm e} \simeq 1.2~n_{\rm p}$.
We apply the same assumptions as \citet{lallement2004}, where $n_{\rm NH}$ and $n_{\rm p}$ to be 0.5 and 2.0, respectively.
From the Chandra X-ray image, we estimate that $r_{s(in)}$ and $r_{s(out)}$ correspond to angular distances of 0.50 and 0.55 arcmin from the center of DEM L71, respectively.

The range of $p$ for region 2 calculated based on Equations (\ref{eq:col_par}) and (\ref{eq:EM_ratio}) is 0.998--0.99992, which means 0.008--0.2 $\%$ of the shock radius is CX dominant.
The emission measure ratio $(EM_{\rm CX}/EM_{\rm ISM})$ and $p$ for all regions are summarized in Table \ref{tab:line_flux}.
Obtained values of CX dominant radius are consistently lower than previous theoretical prediction value of $1~\%$ \citep{lallement2004}.
However previous observations consistently imply smaller than the theoretical prediction, such as 0.2--0.3\% \citep{tanaka2022}.
We therefore conclude that CX is also one of the processes that can explain the observed line ratios within a reasonable parameter range.
If the high $f/r$ ratios are solely attributed to CX, this implies that CX occurs within a more localized region than the previous theoretical prediction \citep{lallement2004}.
To constrain the physical conditions in which CX efficiently take place, a detailed examination of the surrounding gas is necessary.



\section{Conclusions} \label{sec:conclusion}
We performed a spatially resolved high-resolution X-ray spectroscopy of DEM L71 using RGS onboard XMM-Newton.
By utilizing the large dispersion angle of the RGS, we successfully resolved individual line emissions and reconstructed their spatial distribution. 
The RGS images of Fe and O emission lines revealed the following: (1) O K-shell lines predominantly originate from the ISM, while Fe emission is primarily associated with the ejecta; (2) the RGS image exhibits broadening along the dispersion axis, likely caused by the expansion of the ISM and ejecta; and (3) there are spatial variations in the observed $f/r$ ratios of O\emissiontype{VII}.
Our region-by-region plasma diagnostics revealed that the $f/r$ ratios of O\emissiontype{VII} in several regions were higher than those expected for NEI plasmas.
The effect of RS can account for the observed O line ratios. 
The observed $f/r$ ratios show a positive correlation with the O\emissiontype{VII} $f$ line fluxes, suggesting that RS contribute to the observed $f/r$ ratio enhancement.
However, region 6 deviates from this trend, and region 5 is unlikely to exhibit an $f/r$ ratio enhancement through RS because of its plasma morphology. 
Meanwhile, CX can also reproduce the observed O line ratios within reasonable parameter ranges.
To clarify which process is occurring efficiently, it is essential to investigate the surrounding gas distribution in detail.
Our results demonstrate that the RGS, along with X-ray calorimeters such as the Resolve aboard the XRISM, continue to serve as an outstanding X-ray imaging spectrometer for diffuse objects.

\begin{ack}
The authors thank Hitomi Suzuki for her help in the observation proposal and Kosuke Ohba for his help in early data analysis.
The authors also thank Dr. Takaaki Tanaka and Dr. Hiroyuki Uchida for fruitful discussions.
The authors deeply appreciate all the XMM-Newton team members. 
This work is supported by JSPS/MEXT Scientific Research grant Nos. JP24K17106 (Y.A.), JP22KJ1047 (Y.O.), JP24K17093 (H.S.), JP22H00158 (H.Y.), and JP23H01211 (H.Y.).
\end{ack}



\section*{Appendix. Summary of the diagnostic results for all regions}
\begin{figure*}
\begin{center}
\includegraphics[width=17.0cm]{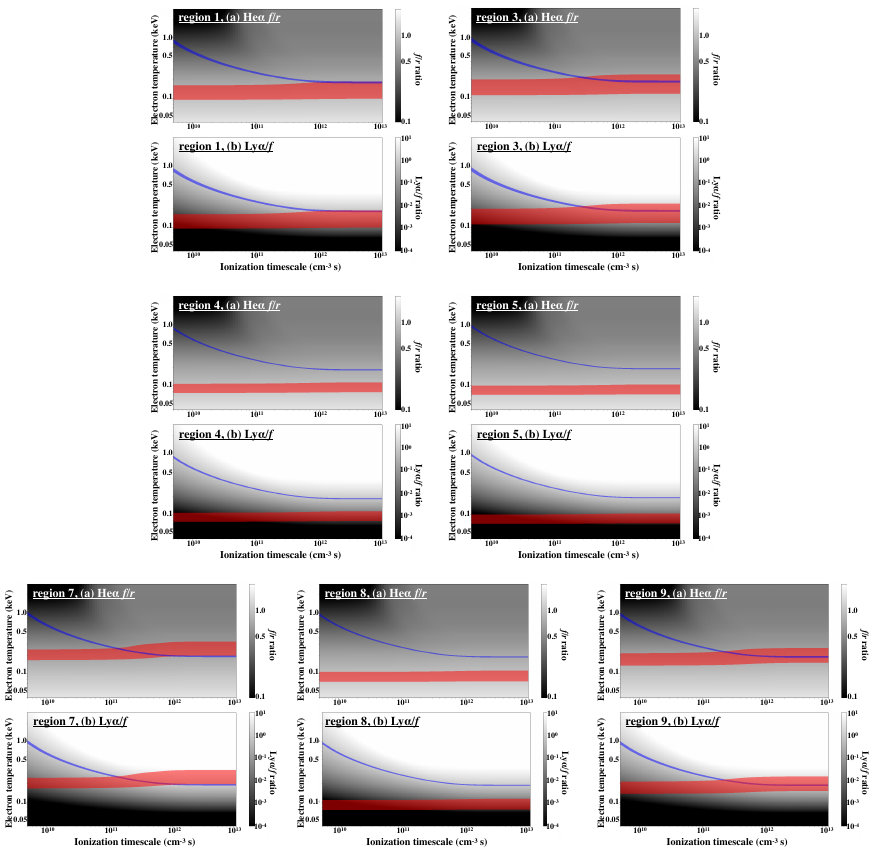} 
\end{center}
\caption{Comparison of the observed O line ratios and plasma models.
The observational values for the regions 1, 3, 4, 5, 7, 8 and 9 are displayed.
{Alt text: Fourteen black and white scale maps.}
}
\label{fig:kTe_net_map_regother}
\end{figure*}

We here present the results of the diagnostics using the collisionally ionized plasma model described in Section \ref{subsec:diagnostics_NEI}.
Figure \ref{fig:kTe_net_map_regother} shows results other than regions 2 and 6.

\bibliography{reference_20241008_3}{}
\bibliographystyle{apj}






\begin{landscape}

\begin{table}[b]
    \centering
    \caption{The line intensities and their ratios, and obtained parameters from charge exchange and resonance scattering estimation.}
    \label{tab:line_flux}
    \renewcommand{\arraystretch}{1.2}
    \begin{tabular}{ccccccccc}
        \toprule
        & \multicolumn{3}{c}{Line flux ($\times 10^{-4}$ photon/s/cm$^{-2}$)} & \multicolumn{2}{c}{Flux ratio} & \multicolumn{2}{c}{CX} & RS \\
        \cmidrule(lr){2-4} \cmidrule(lr){5-6} \cmidrule(lr){7-8} \cmidrule(lr){9-9}
        Region No. & He$\alpha$ $r$ & He$\alpha$ $f$ & Ly$\alpha$ & $f/r$ & Ly$\alpha$/He$\alpha$ $f$ & $EM_{\text{CX}}/EM_{\text{ISM}}$ & $p$ & column density (cm$^{-2}$) \\
        \midrule
        1 & 6.54 $\pm$ 0.56 & 5.71 $\pm$ 0.61 & 4.48 $\pm$ 0.41 & 0.873 $\pm$ 0.119 & 1.01 $\pm$ 0.11 & $\cdots$ & $\cdots$ & $\cdots$ \\
        2 & 8.70 $\pm$ 0.54 & 11.5 $\pm$ 0.6 & 8.42 $\pm$ 0.43 & 1.32 $\pm$ 0.11 & 0.912 $\pm$ 0.055 & 0.0002--0.002 &  0.9992--0.99992 & ($5$--$50) \times 10^{19}$ \\
        3 & 6.78 $\pm$ 0.63 & 5.39 $\pm$ 0.61 & 4.49 $\pm$ 0.44 & 0.796 $\pm$ 0.117 & 0.926 $\pm$ 0.126 & $\cdots$ & $\cdots$ & $\cdots$ \\
        4 & 10.1 $\pm$ 0.6 & 10.2 $\pm$ 0.6 & 8.07 $\pm$ 0.41 & 1.01 $\pm$ 0.08 & 0.791 $\pm$ 0.062 & 0.00005--0.001 & 0.9992--0.99998 & ($2$--$30) \times 10^{19}$ \\
        5 & 10.3 $\pm$ 0.5 & 10.8 $\pm$ 0.6 & 10.1 $\pm$ 0.4 & 1.04 $\pm$ 0.08 & 0.936 $\pm$ 0.068 & 0.00006--0.001 & 0.9992--0.99998 & ($3$--$40) \times 10^{19}$ \\
        6 & 13.8 $\pm$ 0.6 & 10.4 $\pm$ 0.6 & 9.18 $\pm$ 0.44 & 0.756 $\pm$ 0.056 & 0.879 $\pm$ 0.067 & $\cdots$ & $\cdots$ & $\cdots$ \\
        7 & 6.27 $\pm$ 0.42 & 4.16 $\pm$ 0.39 & 4.18 $\pm$ 0.26 & 0.664 $\pm$ 0.077 & 0.785 $\pm$ 0.113 & $\cdots$ & $\cdots$ & $\cdots$ \\
        8 & 7.95 $\pm$ 0.46 & 8.10 $\pm$ 0.52 & 7.38 $\pm$ 0.31 & 1.02 $\pm$ 0.09 & 0.734 $\pm$ 0.071 & 0.00005--0.001 & 0.9992--0.99998 & ($2$--$30) \times 10^{19}$ \\
        9 & 5.93 $\pm$ 0.41 & 4.35 $\pm$ 0.40 & 4.03 $\pm$ 0.27 & 0.733 $\pm$ 0.084 & 0.832 $\pm$ 0.105 & $\cdots$ & $\cdots$ & $\cdots$ \\
        \bottomrule
    \end{tabular}
\end{table}

\end{landscape}

\end{document}